\input harvmac

\Title{\vbox{\baselineskip12pt
\hbox{BCCUNY-HEP/00-01}
\hbox{hep-th/0006063}}}
{\vbox{\centerline{Entropy and String/Black Hole Correspondence}}}

\baselineskip=12pt
\centerline {Ramzi R. Khuri\footnote{$^*$}{e-mail: 
khuri@gursey.baruch.cuny.edu. Supported by NSF Grant 9900773 
and by PSC-CUNY Award 669663.}} 
\medskip
\centerline{\sl Department of Natural Sciences, Baruch College, CUNY}
\centerline{\sl 17 Lexington Avenue, New York, NY 10010
\footnote{$^\dagger$}{Permanent address.}}
\medskip
\centerline{\sl Graduate School and University Center, CUNY}
\centerline{\sl 365 5th Avenue, New York, NY 10036}
\medskip
\centerline{\sl Center for Advanced Mathematical Sciences}
\centerline{\sl American University of Beirut, Beirut, Lebanon
\footnote{$^{**}$}{Associate member.}}

\bigskip
\centerline{\bf Abstract}
\medskip
\baselineskip = 20pt 
We make some observations regarding string/black hole correspondence
with a view to understanding the nature of the quantum degrees of
freedom of a black hole in string theory. In particular, we compare
entropy change in analogous string and black hole processes in order 
to support the intepretation of the area law entropy as arising from 
stringy constituents.

\Date{June 2000}

\def\({\left (}
\def\){\right )}
\def\[{\left [}
\def\]{\right ]}

\lref\thorn{See C. B. Thorn, hep-th/9607204 and references therein; see also
O. Bergman and C. B. Thorn, Nucl. Phys. {\bf B502} (1997) 309.}

\lref\bekhawk {J. Bekenstein, Lett. Nuov. Cimento {\bf 4} (1972) 737;
Phys. Rev. {\bf D7} (1973) 2333; Phys. Rev. {\bf D9} (1974) 3292;
S. W. Hawking, Nature {\bf 248} (1974) 30; Comm. Math. 
Phys. {\bf 43} (1975) 199.}
 
\lref\GSW {M. B. Green, J. H. Schwarz and E. Witten,
{\it Superstring Theory}, Cambridge University Press, Cambridge (1987).}

\lref\prep{See M. J. Duff, R. R. Khuri and J. X. Lu, Phys. Rep.
{\bf B259} (1995) 213, M. Cvetic and D. Youm, Phys. Rev. {\bf D54}
(1996) 2612, M. Cvetic and A. A. Tseytlin, Nucl. Phys.
{\bf B477} (1996) 499 and references therein.} 

\lref\cvet{M. Cvetic and D. Youm, Phys. Rev. Lett. {\bf 75} (1995) 4165.}  

\lref\rahm{J. Rahmfeld, Phys. Lett. {\bf B372 } (1996) 198.}

\lref\pol{J. Polchinski hep-th/9611050 and references therein.}

\lref\stva {A. Strominger and C. Vafa, Phys. Lett. {\bf B379}
(1996) 99.}

\lref\malda{J. Maldacena, hep-th/9607235 and references therein.}

\lref\sfet{K. Sfetsos and K. Skenderis, Nucl. Phys. {\bf B517} (198) 179;
R. Argurio. F. Englert and L. Houart, Phys. Lett. {\bf B426} (1998) 275.}

\lref\holo{G. 't Hooft, gr-qc/9310026; L. Susskind, L. Thorlacius and J. Uglum, 
Phys. Rev. {\bf D48} (1993) 3743.}

\lref\corr{L. Susskind, hep-th/9309145.}

\lref\horpol{G. T. Horowitz and J. Polchinski, 
Phys. Rev. {\bf D55} (1997) 6189.}

\lref\self {G. T. Horowitz and J. Polchinski, Phys. Rev. {\bf D57}
(1998) 2557.}

\lref\random{P. Salomonson and B. S. Skagerstam, Nucl. Phys.
{\bf B268} (1986) 349; Physica {\bf A158} (1989) 499;
D. Mitchell and N. Turok, Phys. Rev. Lett. {\bf 58} (1987) 1577;
Nucl. Phys. {\bf B294} (1987) 1138.}

\lref\polytext{See M. Doi and S. F. Edwards, {\it The Theory of 
Polymer Dynamics}, Clarendon Press, Oxford (1986) and references
therein.}

\lref\poly{S. F. Edwards and M. Muthukumar, J. Chem. Phys. {\bf 89}
(1988) 2435; S. F. Edwards and Y. Chen, J. Phys. {\bf A21}
(1988) 2963.}

\lref\callone{D. J. E. Callaway, Phys. Rev. {\bf E53} (1996) 3738.}

\lref\calltwo{D. J. E. Callaway, PROTEINS: Structure, Function
and Genetics {\bf 20} (1994) 124.}

\lref\apostol{T. M. Apostol, {\it Introduction to Analytic Number
Theory}, Springer Verlag (1976).}

\lref\essay{R. R. Khuri, Mod. Phys. Lett. {\bf A13} (1998) 1407.}

\lref\polk{R. R. Khuri, Phys. Lett. {\bf B470} (1999) 73.}

\lref\damven{T. Damour and G. Veneziano, hep-th/9907030.}

\lref\rk{See R. R. Khuri, hep-th/9609094 and references therein.}

\lref\gps{D. J. Gross and M. J. Perry, Nucl. Phys. {\bf B226} (1983) 29;
R. D. Sorkin, Phys. Rev. Lett. {\bf 51} (1983) 87.}

\lref\hmono{R. R. Khuri, Nucl. Phys. {\bf B387} (1992) 315.}



\newsec{Introduction}

The conjecture of string-black hole correspondence \corr,
in its most basic form, proposes a one-to-one mapping between fundamental string states 
and quantum black hole states. Since it attempts to explain the underlying physics of 
inherently general-relativistic objects such as
black holes in terms of quantum states of string theory, its implications have profound 
consequences for the understanding quantum gravity from string theory.

This correspondence was clarified recently in \horpol, where it was
argued that by adiabatically increasing the string coupling constant
$g$ an excited string state would turn into a Schwarzschild black hole at 
$g=g_c \sim N^{-1/4}$, where $N >> 1$ is the level number of a long, fundamental 
string state. The string coupling takes this critical value precisely when the 
Schwarzschild radius, $R_S$, becomes of the order of the string scale, $l_s$. 
For couplings below $g_c$, the picture 
of a string state prevails, while for coupling above $g_c$ the black hole
picture prevails. At this critical coupling, the string entropy
makes a smooth transition into the Bekenstein-Hawking black hole area
entropy law \bekhawk.

The authors of \horpol\ took this correspondence a step further in
\self, using a thermal scalar field theory formalism to study the 
size of the string state as it collapses from its initial random walk
\random\ form into a black hole. Their analysis in $D=4$ dimensions 
predicted a specific dependence of the size of the string state 
on the string coupling in an intermediate region between the coupling 
$g_0$ at which gravitational effects first become significant,
and the transitional coupling $g_c$. It was then argued in \essay\
(and shown explicitly in \polk) that these results could be derived from methods 
of polymer physics,  with the ``string bit" representing a single step in a random 
walk picture \thorn. The results of \horpol\ were also reproduced in \damven, 
where methods of fundamental string theory were used more directly, and where 
some of the physical issues relating to the collapsing string were clarified. 

It was also argued in \essay\ that string/black hole correspondence
in the case of extremal Reisnner-Nordstrom (RN) solutions of string theory
has a particularly simple realization in terms of the combinatorics of 
constituents, both in the string and in the black hole pictures.
In \polk, the implications of correspondence for Schwarzschild black holes 
were interpreted to explain the transition from the random walk, stringy 
degrees of freedom to the holographic, horizon degrees of freedom.

In this paper, we make some further observations regarding 
this correspondence, and propose the possibility of understanding
quantum gravitational degrees of freedom directly from string theory.
In particular, we compare entropy change in black hole dynamics with analogous processes
for corresponding BPS states in string theory. We first elaborate on the argument 
in \essay\ of ``constituent correspondence'' for extremal black holes and then summarize 
the arguments in \polk\ on the projection of the random walk, stringy degrees of freedom 
onto the horizon, in support of the holographic principle \holo. Combining these results
leads to a simple physical picture for entropy enhancement in black hole processes
in terms of stringy degrees of freedom. We focus throughout on the case of four dimensions, 
with the expectation that our arguments are valid for $D\neq 4$.

\newsec{Constituent Correspondence}

At the heart of the success (first realized in \stva) of string theory in reproducing 
the Bekenstein-Hawking black hole entropy formula \bekhawk\ is the feature of 
compositeness \rk, namely the construction of general solutions as composites 
(or bound states) of fundamental, constituent solutions. These latter represent 
single-charged solutions corresponding to single states in the perturbative or 
nonpertubative spectrum.

The four-dimensional Reissner-Nordstrom solution, for example, arises as a 
supersymmetric solution composed of four fundamental charges arising from
a ten-dimensional string theory (or an eleven-dimensional M-theory). This
composite picture may arise in numerous ways, with or without D-branes.
We adopt the viewpoint that the different ways of obtaining a solution
represent U-dual pictures which are essentially equivalent. In any case,
the common feature arises that the quantum entropy obtained by counting
the degeneracy of string states in the zero coupling limit precisely matches
the area law entropy (the Bekenstein-Hawking law). Supersymmetry nonrenormalization 
theorems are then invoked to justify this agreement by
forbidding quantum corrections to the entropy in increasing the coupling from zero 
to the black hole transition point. The success
of the entropy matching is an example of the string/black hole correspondence 
working perfectly. Not only the functional form, but the precise value, 
of the black hole entropy is recovered in this way.

In \essay, it was argued that correspondence should be taken even more seriously 
in this case: not only are the numbers of degrees of freedom 
equal in the string and black hole pictures, but their nature is essentially
unchanged. In other words the identical combinatorics should be used to count states 
whether in the string or in the black hole picture. 

Let us elaborate a bit on this idea, and consider for simplicity
the solutions without D-branes (ie with NSNS charge only). The
``factorizability" \cvet\ of the general solutions has been seen in many different 
contexts. 
We follow the setup in \rahm\
and consider a four-dimensional solution that arises from 
four consitituent charges: two Kaluza-Klein fields (from the metric) 
and two winding modes (from the antisymmetric tensor), each pair consisting
of one electric and one magnetic charge. Each field on its own would give
rise to an electric or magnetic Kaluza-Klein \gps\
or H-monopole solution \hmono. In the compactification from ten to four dimensions, 
only two of the compactified dimensions result in charges
(each providing an electric/magnetic pair) appearing in four dimensions. 
The remaining four compactifield dimensions remain passive, but their existence is 
nevertheless crucial for the entropy matching.

Let $F_1$ and $F_2$ represent the field
strengths of the electric and magnetic KK fields, respectively, with 
charges $Q_1$ and $Q_2$,
and $F_3$ and $F_4$ represent the field strengths of the electric and magnetic 
H-monopoles, respectively, with charges $Q_3$ and $Q_4$.
Then the four-dimensional metric in the canonical (Einstein) frame
may be written as
\eqn\metric{ds^2 = -\left(H_1 H_2 H_3 H_4\right)^{-1/2} dt^2
+ \left(H_1 H_2 H_3 H_4\right)^{1/2}\left(dx_1^2 + dx_2^2 + dx_3^2\right),}
where $H_k = 1 + {|Q_k|\over |\vec x - \vec y_k|}$, $k = 1, 2, 3, 4$,
and where $\vec y_k$ are the locations of the four constituents in the
three-dimensional space. Note that a horizon with nonzero area can only
be realized provided the four constituents are all placed at the origin
($\vec y_k = 0$, for $k = 1,2,3,4$).
Note also that the existence of this solution is based on an underlying
zero-force condition that allows for the multi-source solutions at
arbitrary locations. Since zero energy is required to displace any of the
constituents, the degeneracy and entropy depend only on the number of 
possible combinations that form a given configuration.

Corresponding to this solution in the black hole picture (valid provided
curvatures are less than $1/\alpha'$), is a collection of states in the
string picture at zero coupling. Corresponding to each charge $Q_k$ is the
eigenvalue of a quantum number operator $N_k$ (momentum for KK, winding for H-monopole). 
The quantum degeneracy $d(N_k)$ of the number of states implies an entropy 
$S_{QM}= \ln d(N_k) = 2\pi \sqrt{|N_1 N_2 N_3 N_4|}$ which is precisely 
equal to the BH area law entropy \malda
\eqn\arealaw{S_{BH} = {A\over 4G} = {4\pi \sqrt{|Q_1 Q_2 Q_3 Q_4|}\over 4G}
= S_{QM}.} 

The RN solution arises provided $Q_1 = Q_2 = Q_3 = Q_4 =Q$. Otherwise,
the solution has in general some nonzero scalar fields. The arguments below 
apply in the more general case, but we focus initially on the RN black hole for simplicity.
Even in the RN solution, the corresponding momentum and winding states 
may have different number operators, as the precise relationship between
the $Q_k$ and the $N_k$ depends on the sizes of the compactified directions.
This last fact therefore gives us the freedom to have the same classical charges but 
different quantum numbers. Nevertheless, the general equality
\arealaw\ of the two entropies persists.

The quantum degeneracy essentially arises from the number of ways of distributing 
$N_1 N_2 N_3 N_4$ states along $4$ bosonic and $4$ fermionic degrees of freedom. 
This yields the above degeneracy and entropy provided the $N_k$ are large. 
What correspondence then tells us is that the same combinatorics should hold 
in the black hole picture as well. In this case, this can be seen directly 
from the form of the solution. Nonzero entropy (or area) can only be realized 
provided none of the four constituent charges
vanish. Another way of saying this is that the horizon forms via string
``nucleations" which occur whenever a unit charge of each species combines
with unit charges from each of the three other species to form a nonzero
horizon ``pixel". These nucleations can occur via bosonic or
fermionic string degrees of freedom condensing along the four degrees of
freedom corresponding to the four ``passive" dimensions 
(which produce no four-dimensional charge upon compactification).
The degeneracy is then given by the number of such nucleations that can occur along
the four bosonic or four fermionic degrees of freedom. 
Hence precisely the same combinatoric picture as in the perturbative picture
arises directly from the solution, 
yielding the same degeneracy and entropy as for the quantum states in the 
zero-coupling limit. Note that this correspondence does not depend on the 
compactification from which the four-dimensional solutions arises.

\newsec{Random Walks and Quantum Degrees of Freedom}

We now consider a long, self-gravitating string at level $N$
and adiabatically increase the coupling $g$ until the string collapses 
into a black hole. As noted in \self, the string size at zero coupling 
(the free string) is initially given by $R_{RW} \sim N^{1/4} l_s$, where 
$l_s$ is the string scale. The letters ``RW" denote ``Random Walk", as the 
free string represents a random walk \random\ with $n=N^{1/2}$ steps (or 
string ``bits" \thorn). The total length of the string is given by $L=nl_s$.
This configuration may be represented as a 
random walk polymer chain with self-interaction. There are $n$ steps,
each of length $l$, with $\vec r_i$ representing the position of the 
chain after the $ith$ step. Gravitational self-interactions start to become 
significant once $g \sim g_0 \sim n^{-3/4}$ \refs{\self,\polk}.

This system is described by the generalized Hamiltonian
\eqn\hamilt{\beta H ={3\over 2l}\int_0^L ds \left({\partial \vec R(s)\over 
\partial s}\right)^2 + g^2l \int_0^L \int_0^L ds ds'{1\over 
|\vec R(s) -\vec R(s')|},}
where $\vec R(s)$ is the position vector of the chain at arc-length
$s$ ($0 \leq s \leq L$). From a Feynman variational
procedure for the free energy of the chain,
It is straightforward to show that the size of the polymer,
the average mean square end-to-end distance of the chain, is given by \polk
\eqn\sizetwo{R^2 \simeq {l_s^2\over g^4 n^2} \left(1 - \exp{(-g^4n^3)}\right).}
For $g << g_0$, $R^2 \sim nl_s^2 $, which is the random walk/free string result, 
while for $g_0 < g < g_c$, $R \sim l_s/(g^2n)$, which agrees with the calculation 
of \refs{\self,\damven}.

At zero coupling, $S_0 \sim n$, the number of steps of the random walk.
In the intermediate range $g_0 < g < g_c$, the adiabatic increase of the
coupling preserves the essential degrees of freedom associated with the
string bits. Of course one no longer has a random walk, but up to a
factor of order unity, the string bits retain most of their degrees of
freedom.

For a black hole whose mass is equal
to the excited string state up to a factor of $O(1)$, 
$S_{BH}=A/4G \sim R_S^2/l_P^2$, where $l_P=g l_s$ is the Planck scale and 
$R_S \sim GM \sim l_P^2 M \sim l_P^2n/l_s \sim g^2n l_s$ 
is the Schwarzschild radius. We may then rewrite the BH entropy as
\eqn\bhent{S_{BH} \sim {n^2 l_P^2\over l_s^2}\sim n^2 g^2.}
At the critical coupling $g_c \sim n^{-1/2}$, the entropy makes a smooth 
transition to the Bekenstein-Hawking area law form: $S_{BH} \sim n \sim S_0$. 
This entropy still represents the degeneracy of a polymer 
system with $n$ steps (or links). This can be seen as follows: at $g=g_c$, the size
of the collapsed polymer string is given by $R \sim R_S \sim l_s$, or the
size of one string bit, or one step. In order for $n$ steps of size $l_s$
to fit into a sphere of radius $R_S$, the number of possible positions to 
which each step can go must remain a small whole number, $p'$. Hence the 
degeneracy again has the random walk form $d \sim p'^n$.
Of course, the polymer is no longer a random walk, but the degrees of 
freedom still remain essentially intact. 

Once the transition is complete and the black hole picture prevails,
the area of the black hole is given by 
$A\sim R_S^2 \sim l_s^2 \sim (1/g^2) l_P^2 \sim n l_P^2$. So the horizon
can be divided into $n$ ``pixels" each of area $l_P^2$. Once the horizon forms,
the degrees of freedom associated with it represent independent quantum
states, the points on the horizon are causally disconnected.
Again, only a small whole number of possible states, 
$q$, is associated with each pixel, so that the total degeneracy is given
by $d_{BH} \sim q^n$, with the entropy given by 
$S = \ln d \sim n \sim S_{BH}$. 
So essentially the random walk degrees of freedom turn into horizon
surface degrees of freedom at the critical transition point. Another way 
of saying this is that the string bits project their information onto the 
horizon, in accord with expectations of the holographic
principle \holo. So the underlying degrees of 
freedom of quantum black holes in string theory remain associated 
with the original, stringy degrees of freedom. This further strengthens
the string/black hole correspondence conjecture \corr\ by implying that
in the transition to the strong-coupling limit, it is possible that the
quantum string states somehow retain far more of their nature from the
perturbative picture than might have been supposed.

\newsec{String/Black Hole Correspondence and Entropy Enhancement}

We now wish to employ both the constituents of section 2 and the random walk
degrees of freedom of section 3 in string/black hole correspondence
to shed light on the underlying quantum degrees of freedom of a black hole in string 
theory. At weak coupling, the random walk entropy is given by $S_{RW} = 2\pi n =
2\pi \sqrt N$, where $n$ and $N$ are the number of steps and level number 
respectively. Consider now the BPS state corresponding to the weak coupling
limit of an extremal RN black hole. This state typically arises as a composite
of four charges, with number operator eigenvalues $N_1, N_2, N_3$ and $N_4$.
The entropy is given by $S = 2\pi \sqrt {|N_1 N_2 N_3 N_4|}$, as indicated above.
In the random walk picture, this represents a composite of four independent random
walks, each with $n_i = \sqrt |N_i|$ steps. For example, if we set $N_2 = N_3 = N_4 =1$,
then $n=n_1 = \sqrt {|N_1|}$ is the number of steps. In the general case,
$n = n_1 n_2 n_3 n_4 = \sqrt{|N_1 N_2 N_3 N_4|} = \sqrt{N}$.

Suppose we combine two such BPS states in the weak coupling limit with 
initial quantum numbers $(N_1, N_2, N_3, N_4)$ and 
$(N_1', N_2', N_3', N_4')$ and entropies 
$S = 2\pi \sqrt {|N_1 N_2 N_3 N_4|}$ and
$S' = 2\pi \sqrt {N_1' N_2' N_3' N_4'}$, respectively.
Assume first that $N_i$ and $N_i'$ have the same sign. The combined state 
has quantum numbers $(N_1 + N_1', N_2 + N_2', N_3 + N_3', N_4 + N_4')$.
From the zero-force condition between such BPS states, the entropy  depends simply on
the number of possible configurations and is given by
\eqn\tentsameone{S_T = 2\pi \sqrt {|(N_1 + N_1') (N_2 + N_2') 
(N_3 + N_3') (N_4 + N_4')|}.}

Now consider the fusion of the corresponding 
extremal four-dimensional RN black holes in the black hole limit. The RN black hole 
carries a single charge in four-dimensions, albeit arising as the composite of four
number operators. Let us represent a general RN black hole with mass
$M$ and charge $Q$ by $(M, Q)$. Henceforth we set $G=1$ for simplicity.
The event horizon radius is given by 
\eqn\rad{R = M + \sqrt {M^2 - Q^2}.} 
An extremal
black hole with positive charge is represented by $(M, M)$, has radius 
$R = M$ and
entropy $S_{BH} = 4\pi R^2/4 \pi  = M^2$. A second extremal, positively charged
black holes may be represented by $(M', M')$, where $M' = \alpha M$
for some positive $\alpha$. 
The initial entropies of the black holes before fusion are given by 
$S_{BH}=M^2$ and $S_{BH}' = M'^2 = \alpha^2 S_{BH}$, so that
the total initial entropy is 
\eqn\totentisamebh{S_{BH}^i = (1+ \alpha^2) S_{BH}.}
As a result of the zero force condition, these two black holes can be 
combined with zero energy into a single extremal 
RN black hole $(M+M', M+M')$ with radius $R = M+M'$ and entropy
\eqn\tentsametwo{S_{BH}^f = (M+M')^2 = (1+\alpha)^2 S_{BH}.}

{}For two interacting four-dimensional RN black holes arising from the same string 
theory, the compactification radii along each direction must be the same. This implies
that $Q_i = k_i N_i$ and $Q'_i = k_i N'_i$ for the same constants $k_i$.
In order to obtain RN black holes, we set $Q_1=Q_2=Q_3=Q_4$ and
$Q'_1=Q'_2=Q'_3=Q'_4$, from which it follows that $N_i' = c N_i$, where
$c >0$ is the same constant for $i=1,2,3$ or $4$. From string/black hole correpondence
and the equality of the individual entropies in the perturbative and black hole limits,
$S=2\pi\sqrt{|N_1 N_2 N_3 N_4|} = S_{BH}$ above and
$S'= 2\pi \sqrt {|N_1' N_2' N_3' N_4'|} = c^2 S = c^2 S_{BH} = S_{BH}' = \alpha^2 S_{BH}$.
It follows that $c = \alpha$, from which we obtain 
\eqn\tentsamethree{S_T = 2\pi \sqrt {(N_1 + N_1') (N_2 + N_2') 
(N_3 + N_3') (N_4 + N_4')} = (1+\alpha)^2 S = S_{BH}^f.}
So the same process is at work both in the random walk and black hole pictures, 
with the same entropy enhancement resulting from simple counting of configurations
of constituents.

Now consider instead the combining of oppositely charged BPS states. Here the
zero-force condition no longer holds and the combined state is not as simply obtained
as in the same charge case. In order to again compare with four-dimensional RN
black holes, we must have uniformly proportional but opposite sign quantum numbers.
For example, if one of the operators represents winding around a compactified direction,
then the second BPS state would possess winding around the same compactified direction but
in the opposite direction. For simplicity,
assume the the first BPS again is represented by $(N_1, N_2, N_3, N_4)$, while
the second is represented by $(-\bar N_1, -\bar N_2, -\bar N_3, -\bar N_4)$. Without
loss of generality, we may assume $N_i, \bar N_i > 0$. 
Then the entropy of each state before they combine is again
given by $S = 2\pi \sqrt {N_1 N_2 N_3 N_4}$ and 
$\bar S  = 2\pi \sqrt {\bar N_1 \bar N_2 \bar N_3 \bar N_4}$. 
Here $\bar n_i = \sqrt {\bar N_i}$ is the number of steps of the random walk 
corresponding to each of the four species. Combining the two states
is somewhat more complicated than in the same charge case above, since we no longer
have a zero-force condition and the entropy is no longer simply configurational.
Here the number
configurations corresponds to the combination of independent random walks for 
each species of quantum operator. However, in this case, rather than simply add
the quantum numbers for each species, we must add the steps for each random walk.
This is simply a consequence of the fact that we effectively have a random walk that
is $n_i + \bar n_i$ steps for $i=1,2,3,4$. A simpler way of viewing this is to start
at the end of one of the random walks and proceed to the end of the other.
We then have to proceed by $n_i + \bar n_i$ steps to
complete all possible configurations corresponding to the combination of the two
states. Therefore the entropy is given by
\eqn\tentoppone{\eqalign{S_T &= 2 \pi (n_1 + \bar n_1)(n_2 + \bar n_2)
(n_3 + \bar n_3)(n_4 + \bar n_4)\cr
&=2\pi (\sqrt N_1 + \sqrt{\bar N_1}) (\sqrt N_2 + \sqrt{\bar N_2})
(\sqrt N_3 + \sqrt{\bar N_3}) (\sqrt N_4 + \sqrt{\bar N_4}).\cr}}

This is a very
different result from the same charge case and should be reproduced in the black
hole limit. Here we represent the positively and negatively charged
extremal black holes as $(M, M)$ and $(M', -M')$, respectively, where 
here $M'/M = \beta >0$. The two black holes before fusion have entropies
$S_{BH} = M^2$ and $\bar S_{BH} = M'^2 = \beta^2 S$, so that the total
initial entropy is $S_{BH}^i = (1+\beta^2) S$. The black hole resulting
from the fusion of these two black holes is of course no longer extremal,
and has mass and charge $(M+M', M-M')$. The radius of the event horizon in this
case is given by 
\eqn\radopp{R = (M+M') + \sqrt{(M+M')^2 - (M-M')^2} = (1+\beta + 2\sqrt\beta)M=
(1+\sqrt\beta)^2 M.}
This implies a total entropy of 
\eqn\tentopptwo{S_{BH}^f = (1+\sqrt\beta)^4 S_{BH}.}  
Of course $S_{BH}^f > S_{BH}^i$, in accord with the second law of black hole thermodynamics,
such a fusion representing an irreversible process. Note that again the entropy enhancement
follows precisely the result for the BPS states. For $\bar N_i = \beta N_i$, \tentoppone\
implies
\eqn\tentoppthree{S_T = (1+\sqrt\beta)^4 S =(1+\sqrt\beta)^4 S_{BH} = S_{BH}^f.} 
The special case of a Schwarzschild black hole arises for $\beta = 1$, in which case
the enhancement $S_{BH}^f/S_{BH}^i = (1+\sqrt\beta)^4/(1+\beta^2) = 8$.
The large increase in entropy is mainly due to the increase of 
gravitationally bound energy at the expense of free energy as a result of the 
elimination of some of the electrostatic force, which previously had balanced 
the gravitational one. 

The above results can be generalized to the non-RN case, in which the four charges
$Q_i$ are not all equal. Suppose we fuse two extremal black holes, 
with constituents of mass and charge $(M_i, Q_i)$ and $(M'_i, Q'_i)$ for $i= 1,2,3,4$.
Here $Q_i = \pm M_i$. Then the black hole obtained from fusion has constituents
$(M_i + M'_i, Q_i + Q'_i)$, $i=1,2,3,4$ with total mass and charge 
\eqn\totmq{\eqalign{\bar M & = \sum_{i=1}^4 (M_i + M'_i) ,\cr
\bar Q & = \sum_{i=1}^4 (Q_i + Q'_i) .\cr}}
Alternatively, one may always
regard the general solution as arising from the fusion of
two extremal black holes.
It is a straightforward exercise to show that the area law entropy of a black hole with 
constituents $(\bar M_i, \bar Q_i)$ is given by
\eqn\genent{\eqalign{S = {\pi\over 4}&
\left(\sqrt{\bar M_1 + \bar Q_1} + \sqrt{\bar M_1 - \bar Q_1}\right)
\left(\sqrt{\bar M_2 + \bar Q_2} + \sqrt{\bar M_2 - \bar Q_2})\right)\cr
&\left(\sqrt{\bar M_3 + \bar Q_3}+ \sqrt{\bar M_3 - \bar Q_3})\right) 
 \left(\sqrt{\bar M_4 + \bar Q_4}+ \sqrt{\bar M_4 - \bar Q_4}\right).\cr}}
For an extremal black hole with $\bar Q_i=\pm \bar M_i$, one obtains the usual formula
\eqn\extent{S = \pi \sqrt{|\bar Q_1 \bar Q_2 \bar Q_3 \bar Q_4|}=
 2\pi \sqrt{|\bar N_1 \bar N_2 \bar N_3 \bar N_4|}.}
For the general solution above regarded as the fusion of two extremal solutions,
the factor contributed by each of the four constituent species to the entropy
depends on whether the charges of the constituents of that species are of the same
or opposite sign. For example, if in the above fusion, $Q_1$ and $Q'_1$
are of the same sign, then $\bar M_1 = M_1 + M'_1$, 
$\bar Q_1 = Q_1 + Q'_1 = \pm \bar M_1$ so that the corresponding factor in the entropy 
\eqn\factorone{\sqrt{\bar M_1 + \bar Q_1} + \sqrt{\bar M_1 - \bar Q_1})=
\sqrt{|2\bar Q_1|}= \sqrt{|2(Q_1 + Q'_1)|},}
corresponding to the factor $\sqrt{|\bar N_1|} = \sqrt{|N_1| + |N'_1|}$ 
in the BPS entropy formula.
By contrast, if, say, $Q_2$ and $Q_2'$ are of opposite sign, 
then $\bar M_2 = M_2 + M'_2$, 
$\bar Q_2 = Q_2 + Q'_2 = \pm (M_2 - M'_2)$ so that the corresponding factor in the entropy 
\eqn\factortwo{\sqrt{\bar M_2 + \bar Q_2} + \sqrt{\bar M_2 - \bar Q_2}
=\sqrt{|2Q_2|}+\sqrt{|2Q'_2|},}
corresponding to the factor $\sqrt{|N_2|} + \sqrt{|N'_2|}$ in the BPS entropy formula.
The difference again follows from the presence or lack of the zero-force condition.
In the same sign case, the entropy enhancement is purely configurational and can be read
from the simple addition of quantum numbers. For the opposite sign case, the combination
of two random walks is the correct interpretation. The entropy formula for the 
general state, here regarded as a fusion of two BPS states with quantum numbers
$(N_1, N_2, N_3, N_4)$ and $(N'_1, N'_2, N'_3, N'_4)$ corresponding to \genent\ is
then given by
\eqn\genentstate{S = 2\pi \prod_{i=1}^4 
\sqrt{|N_i| + |N'_i| + \left(1 - {N_i N'_i \over |N_i N'_i|}\right) |N_i N'_i|}.}

Note that the above results are not new: we obtained the same formula as that found
for near-extremal black holes \malda. Also, the correspondence between the
Schwarzschild entropy and D-brane configurations has already been made \sfet. What is
interesting, however, is that the near-extremal entropy formula appears without
correction explicitly in this case. Also, the physical basis for the correspondence,
in terms of constituents and random walk steps, is illuminated. Nevertheless, these
results require further clarification in order to understand the underlying physics
of how precisely quantum black hole states arise from string theory. In this regard, the
stringy approach of \damven\ may shed light on this process. It is also interesting to 
speculate on whether such a simple correspondence can point to a more directly 
string-theoretic formulation of general relativity, in which the laws of black hole 
thermodynamics arise directly in GR from a counting of string states, without the need 
to invoke supersymmetry. Finally, the robustness of string/black hole correspondence
very likely points to a direct resolution of the black hole information paradox using
methods of string theory.

\listrefs
\end